\documentclass[10pt,twocolumn,letterpaper]{article}

\usepackage{iccv}
\usepackage{times}
\usepackage{epsfig}
\usepackage{graphicx}
\usepackage{amsmath}
\usepackage{amssymb}

\usepackage[breaklinks=true,bookmarks=false]{hyperref}

\iccvfinalcopy 

\def\iccvPaperID{****} 

\ificcvfinal\fi

\begin{document}

\title{Prostate Cancer Inference via Weakly-supervised Learning using a Large Collection of Negative MRI}

\author{
Ruiming Cao\\
University of California, Los Angeles, USA\\
{\tt\small caoruiming@gmail.com}\\
\and
Xinran Zhong\\
University of California, Los Angeles, USA\\
{\tt\small zhongxr9@gmail.com}\\
\and
Fabien Scalzo\\
University of California, Los Angeles, USA\\
{\tt\small fscalzo@mednet.ucla.edu}\\
\and
Steven Raman\\
University of California, Los Angeles, USA\\
{\tt\small SRaman@mednet.ucla.edu}\\
\and
Kyunghyun Sung\\
University of California, Los Angeles, USA\\
{\tt\small ksung@mednet.ucla.edu}\\
}

\maketitle
\ificcvfinal\thispagestyle{empty}\fi

\begin{abstract}
Recent advances in medical imaging techniques have led to significant improvements in the management of prostate cancer (PCa). In particular, multi-parametric MRI (mp-MRI) continues to gain clinical acceptance as the preferred imaging technique for non-invasive detection and grading of PCa.
However, the machine learning-based diagnosis systems for PCa are often constrained by the limited access to accurate lesion ground truth annotations for training. 
The performance of the machine learning system is highly dependable on both quality and quantity of lesion annotations associated with histopathologic findings, resulting in limited scalability and clinical validation.
Here, we propose the baseline MRI model to alternatively learn the appearance of mp-MRI using radiology-confirmed negative MRI cases via weakly supervised learning.
Since PCa lesions are case-specific and highly heterogeneous, 
it is assumed to be challenging to synthesize PCa lesions using the baseline MRI model, while it would be relatively easier to synthesize the normal appearance in mp-MRI.
We then utilize the baseline MRI model to infer the pixel-wise suspiciousness of PCa 
by comparing the original and synthesized MRI with two distance functions.
We trained and validated the baseline MRI model using 1,145 negative prostate mp-MRI scans. 
For evaluation, we used separated 232 mp-MRI scans, consisting of both positive and negative MRI cases.
The 116 positive MRI scans were annotated by radiologists, confirmed with post-surgical whole-gland specimens.
The suspiciousness map was evaluated by receiver operating characteristic (ROC) analysis for PCa lesions versus non-PCa regions classification and free-response receiver operating characteristic (FROC) analysis for PCa localization.
Our proposed method achieved 0.84 area under the ROC curve and 77.0\% sensitivity at one false positive per patient in FROC analysis. 
\end{abstract}

\section{Introduction}
Prostate cancer (PCa) is one of the most common cancer-related diseases among men in the United States~\cite{Siegel2019Cancer2019.}. 
Recent advances in medical imaging have led to significant improvements in the management of PCa. 
In particular, multi-parametric magnetic resonance imaging (mp-MRI) continues to gain clinical acceptance as the preferred imaging technique for non-invasive detection and grading of PCa. 
However, the current standardized image acquisition and reporting structure of prostate MRI, such as Prostate Imaging - Reporting and Data System version 2 (PI-RADSv2), has limited ability to accurately distinguish between indolent and clinically significant PCa
due to its qualitative or semi-quantitative assessment of the imaging~\cite{weinreb2016pi}.
As a result, there often exists over-detection of indolent PCa and under-detection of csPCa, which lead to detrimental overtreatment and undertreatment. Consequently, there is an urgent clinical need to achieve accurate detection and classification of csPCa.

Recent studies have explored quantitative interpretations of mp-MRI by training machine learning models~\cite{litjens2014computer,song2018computer,tsehay2017biopsy,wang2018automated,fehr2015automatic}. 
Most of the machine models were trained under strong supervision using the lesion annotations as the ground truth, 
and thus the performance of the models is dependent on both quantity and quality of training data associated with ground truth annotations.
However, the radiologic findings from mp-MRI are not easy to be fully integrated with histologic findings due to misregistration or insufficient histologic information, resulting in a limited number or quality of ground truth annotations available. 
Litjens \textit{et al.} used MR-guided biopsy dataset to identify biopsy-confirmed lesions in MRI~\cite{litjens2014computer}, and Fehr \textit{et al.} annotated PCa region of interest (ROI) using post-surgical whole-gland specimens as a reference~\cite{fehr2015automatic}. 
Nevertheless, both studies used relatively small numbers of cases (348 and 147 cases, respectively) due to the limited availability of ground truth annotations.

In contrast, the number of mp-MRI scans has been increased in recent years as mp-MRI gains clinical acceptance for a non-invasive diagnostic tool for detecting and grading PCa. 
Many of mp-MRI scans are sometimes ruled out to be MRI negative, showing no visible MRI lesions. The negative MRI case is shown to be reliable without the need for histologic confirmations~\cite{Hamoen2015UseMp-MRI}. 
Thus, the collection of negative MRI scans is more plausible to access in a large quantity than the collection of positive MRI scans with accurate lesion annotations.
While the negative MRI scans are vastly available, the existing machine learning models for detection of PCa cannot solely learn from the negative MRI scans since they need to be trained under strong supervision between normal and PCa lesions. 

In this work, we first propose the baseline MRI model that learns the general appearance of prostate MRI from the negative MRI scans. The baseline MRI model is implemented as a convolutional neural network (CNN) to synthesize a partially-obstructed region of a prostate MR image using the rest of the unobstructed image as the input via weakly-supervised learning. 
Since PCa lesions are case-specific and highly heterogeneous, it would be difficult to synthesize PCa lesions when the baseline MRI model is trained with only negative MRI scans while it is relatively easy to synthesize the normal appearance of prostate MRI. 
Based on this assumption, we use the trained baseline MRI model to infer the cancer suspiciousness map.
Given a testing image set that contains both negative and positive MRI, 
the baseline MRI model synthesizes for different regions from the collection of regions of interest (ROIs), 
and the cancer suspiciousness map is summarized by comparing the original image regions and the synthesized image regions. 

We summarize our contributions as follows. 
We proposed the cancer inference that utilize the baseline MRI model to predict pixel-wise levels of overall suspiciousness via weakly supervised learning, without the need for PCa annotations during training. 
We trained the baseline MRI model using 1,145 negative mp-MRI scans, identified from 3,127 total collected mp-MRI scans from 2016 to 2018 at a single institution. 
We evaluated the proposed cancer inference in a separate testing dataset, 
consisting of highly curated 116 positive, confirmed with histologic whole-gland specimens, 
and 116 negative mp-MRI scans.

\section{Materials and methods}

\subsection{Negative prostate MRI}
\label{sec_negscans}
With IRB approval, we collected 3,127 3 Tesla (3T) prostate mp-MRI scans from 2016 to 2018 at a single institution. 
MRI scans with following conditions were excluded: 1) patients scanned with non-3T MRI scanners, 2) patients scanned with an endorectal coil, 3) patients scanned immediately after prostate biopsy, and 4) patients underwent focal therapy and/or partial prostatectomy. Clinical radiology reports associated with mp-MRI were used to identify negative MRI cases.
We parsed the plain text in the report, reviewed by genitourinary (GU) radiologists following the standardized interpretation guideline of PI-RADSv2~\cite{weinreb2016pi}, into a structured format and identify MRI-negative cases based on two criteria: 1) “no suspicious target was seen” in Finding section, and 2) “no more than mildly suspicious finding” in Impression section. 
We manually examined a random subset to ensure the correctness of the identified negative MRI cases. 
A total of 1,261 negative MRI scans were identified, and we divided them into training, validation, and testing sets, containing 1,095, 50, and 116 cases, respectively.

For each scan, we used the axial turbo spin-echo (TSE) T2-weighted (T2w) (TR/TE, 3800-5040/101ms; FOV, 14$\times$14cm$^2$; matrix, 256$\times$205; slice thickness, 3 mm; no gap) and maps of apparent diffusion coefficient (ADC) using single-shot echo-planar imaging (SS-EPI) DWI (TR/TE, 4800/80ms; FOV, 21$\times$26cm$^2$; matrix, 130$\times$160; slice thickness, 3.6 mm; b-values, 0/100/400/800 s/mm$^2$). 
ADC was registered into T2w, with 0.625$\times$0.625mm$^2$ in-plane resolution and 3mm through-plane resolution. 
Both T2w and ADC were cropped into a small field-of-view (8$\times$8cm$^2$) to improve the model convergence.  
Four consecutive slices around mid and base gland were selected for each scan, resulting in a total of 4,380 slices for training.

\begin{figure}[!t]
\begin{center}
\includegraphics[width=3in]{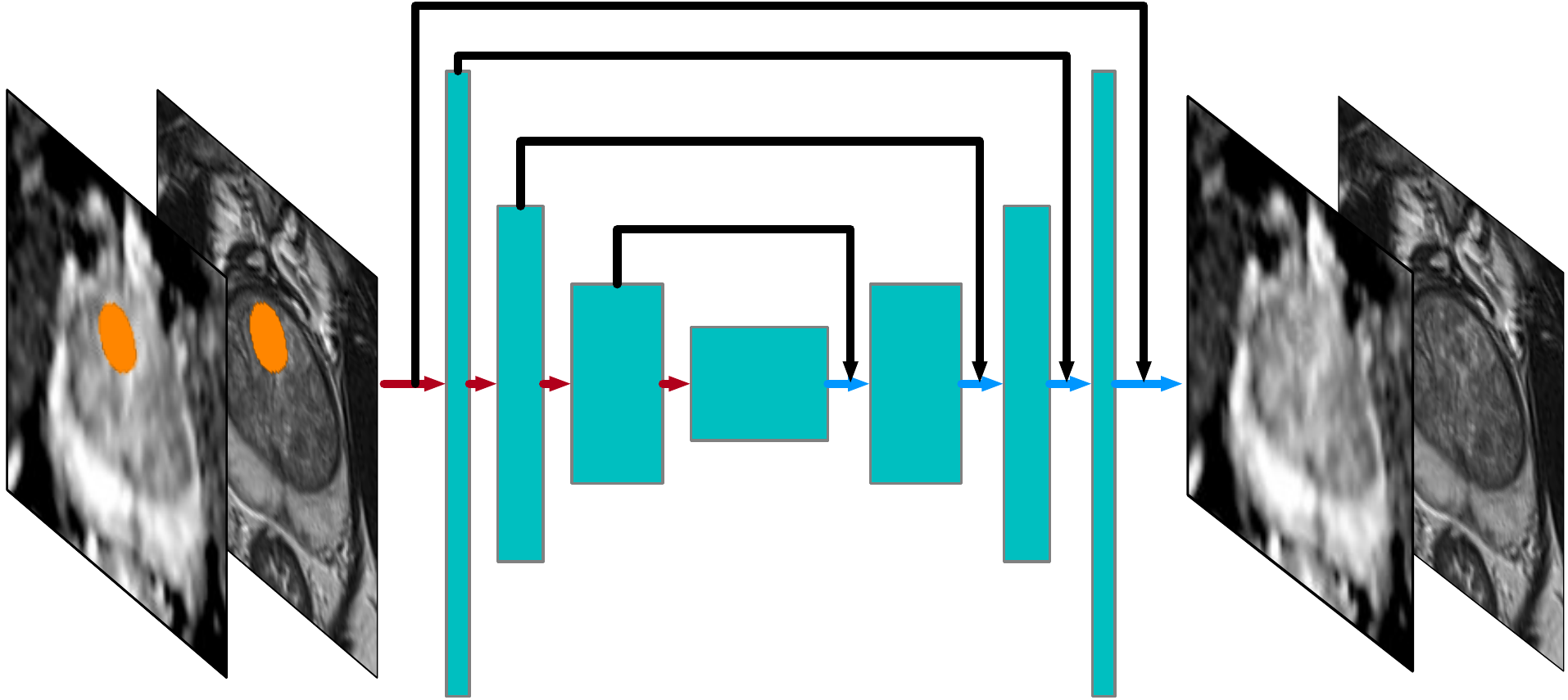} 
\end{center}
\caption{The baseline MRI model synthesizes for the partially obstructed region $M$ (shown in orange) using an unobstructed image as input.}
\label{fig_baselineMRI}
\end{figure}

\subsection{Baseline MRI model}
The baseline MRI model, $f$, aims to recover a partially-obstructed region, defined as a binary mask $M$, by synthesizing similar appearance using the unobstructed image,$\left( 1 - M \right) I$, as input for a given image, $I$.
When we stack T2w ($I_{T2w}$) and ADC ($I_{ADC}$) as different imaging channels ($I=(I_{T2w},I_{ADC})$), 
the baseline MRI model synthesizes, as shown in \figurename~\ref{fig_baselineMRI}, 
the specific region of the stacked image by,
\begin{equation}
M f \big( \left( 1 - M \right) I ; \theta \big) \rightarrow MI, 
\end{equation}
where $\theta$ is the trainable weights of the baseline MRI model. 
We trained the baseline MRI model using only negative MRI scans
so that the baseline MRI model learns the various normal prostate appearance of mp-MRI in training.

We used a U-Net CNN structure~\cite{ronneberger2015u} for the baseline MRI model since the encoder-decoder design of U-Net helps to summarize the global anatomical information~\cite{Dalca2018AnatomicalSegmentation}, 
and the skip connections from U-Net simplify the training for observed regions. 
 The unmasked input was fed directly into the last decoding layer without a need to learn through the encoder-decoder.
In addition, we used partial convolutional layers instead of full convolutional layers to compensate for the zeroed-out input region during encoding~\cite{Liu2018ImageConvolutions}.
We operated the baseline MRI model with 2D images due to the non-isotropic resolution of mp-MRI.

\begin{figure}[!t]
\begin{center}
\includegraphics[width=1.5in]{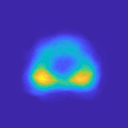} 
\end{center}
\caption{The prevalence map, $P$, constructed by the collection of ROI candidates, accounting for the common locations and shapes of prostate cancer.}
\label{fig_preva_map}
\end{figure}

A collection of ROI candidates, described by the common locations and shapes of PCa, 
was also constructed to avoid learning from irrelevant areas in the image 
(e.g., muscle, fat, bone, etc). 
A total of 1,055 2D ROIs was used from a separate study cohort without any case overlapping~\cite{Johnson2018DetectionImaging}. 
For each 2D ROI, the in-plane location relative to the center of the prostate was maintained, and the through-plane position was ignored. 
Each ROI was converted into a binary mask for the baseline MRI model as an ROI candidate to specify a region $M$ to synthesize. 
As all the ROI candidates were considered in one plane, the collection of ROI candidates, $\mathcal{M}$, accounted for the common locations and shapes of PCa.
A prevalence map, $P$,  constructed by $P=\sum_{M\in\mathcal{M}}M$ is shown in \figurename~\ref{fig_preva_map}. 

We trained the baseline MRI model using the combination of L1 loss, perceptual loss, and style loss~\cite{Gatys2016ImageNetworks}.
The VGG-19 network pre-trained for image classification is used for the calculation of perceptual loss and style loss. 
We only take the feature map from the first convolutional layer for perceptual loss and style loss, since the network is trained for natural images and the higher-level features are not applicable to our context. 
The same weighting for loss terms is used as in~\cite{Liu2018ImageConvolutions}.
The baseline MRI model was trained for 4,000 epochs using a mini-batch of eight $128 \times 128$ training images. 
The learning rate was set to 0.0002 in first 1,000 epochs and was reduced to 0.00005 in the remaining 3,000 epochs with the batch normalization for the encoder turned off as suggested in~\cite{Liu2018ImageConvolutions}. 
Common image augmentations, including shifting, left-right flipping, and gray value variations~\cite{ronneberger2015u}, were applied. 
We also randomly combined multiple ROI candidates together to accelerate training. 
The training took two days using one NVIDIA Titan Xp GPU. 

\begin{figure*}[t!]
\begin{center}
\includegraphics[width=\textwidth]{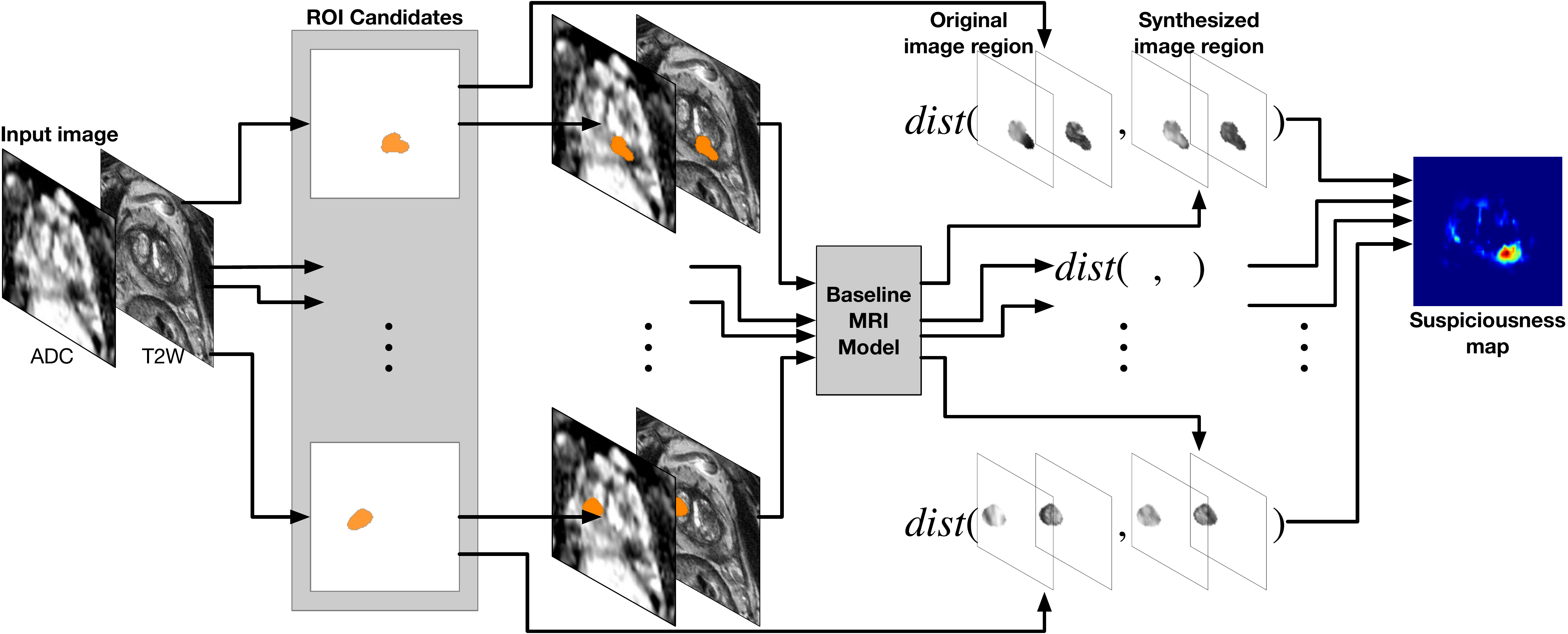}
\end{center}
\caption{The inference of the PCa suspiciousness map using the trained baseline MRI model given an input testing image. The baseline MRI model synthesizes regions specified from the collection of ROI candidates. $dist$ is the distance function for the original image region and the synthesized image region.}
\label{fig_method}
\end{figure*}

\subsection{Cancer inference via baseline MRI model}
\label{sec_susp_inference}
The baseline MRI model was utilized to predict pixel-wise levels of overall suspiciousness for a given testing image. 
The baseline MRI model is assumed to synthesize better when negative MRI is partially obstructed than when positive MRI is partially obstructed since it was trained by only negative MRI cases. 
The regions were considered to be highly suspicious when the difference between the original and synthesized regions is nontrivial. 
In each time, we specify a region to synthesize from the collection of ROI candidates, $M\in\mathcal{M}$, and the synthesized image region from the baseline MRI model is $M f\left( \left( 1 - M \right) I^t;\theta \right)$ where $I^t=\left(I_\mathrm{T2w}^t, I_\mathrm{ADC}^t\right)$ is the testing image. 
By synthesizing different image regions with different ROI candidates, we can obtain the suspiciousness map by
\begin{equation}
    Susp \left( I^t \right) = \frac{1}{P}\sum_{M\in\mathcal{M}} dist \left( M I^t, M f \left( \left( 1 - M \right) I^t; \theta \right) \right) ,
\end{equation}
where $dist\left( I^{ori}, I^{syn} \right)$ is the distance function measuring the pixel-wise difference between the original image region and the synthesized image region, and $P$ is the prevalence map to normalize the suspiciousness map. 
\figurename~\ref{fig_method} illustrates the proposed cancer inference utilizing the baseline MRI model via weakly supervised learning.

Two common distance functions were tested individually: \textit{T2w SSIM} and \textit{ADC Increment},
where \textit{SSIM} is the structural similarity, and \textit{Increment} is the signal intensity increment of the synthesized region compared with the original. 
We evaluated the variation of T2w by \textit{T2w SSIM}, s.t., $dist \left( I^{ori}, I^{syn} \right) = 1 - SSIM \big( I^{ori}_\mathrm{T2w}, I^{syn}_\mathrm{T2w}\big)$, 
since T2w typically contains structural information. 
We measured the ADC intensity increment of the synthesized region compared with the original region by $dist \left( I^{ori}, I^{syn} \right) =  \max \big( I^{syn}_\mathrm{ADC} - I^{ori}_\mathrm{ADC}, 0 \big)$ since ADC is quantitative imaging, and PCa lesion usually has lower ADC intensity than normal tissues~\cite{peng2013quantitative}. 
The suspicion for PCa is high if the ADC intensity in the original region is lower than in the synthesized negative MRI region. 

\section{Experiments}
\subsection{Evaluation dataset}
\label{sec_evaldata}
A separate independent dataset was used for testing the cancer inference, which consisted of 116 positive and 116 negative mp-MRI cases.
For positive MRI, we included pre-operative 3T mp-MRI scans prior to robotic-assisted laparoscopic prostatectomy from 2013 to 2015.
Patients with prior radiotherapy or hormonal therapy were not included.
The grountruth annotations for positive MRI cases were done by GU radiologists who retrospectively reviewed mp-MRI, referring to whole-gland surgical specimens and pathology reports. 
Each MRI visible lesion was matched to the corresponding location on 
the prostate specimen through visual co-registration. 
Later, clinical research fellows, supervised by GU radiologists, annotated all MRI-visible PCa lesions (Gleason Score$\geq$3+4).
We also included prospectively missed PCa lesions (false positives) that are visible in mp-MRI in a retrospective review, but MRI non-visible missed PCa lesions were not included in the study.
The negative MRI cases (116 out of 1,261) were from the same negative prostate MRI dataset pool, described in ~\ref{sec_negscans}.
The ground truth ROIs were annotated on T2w, and the FOV and slice were determined in the same way. 

\begin{figure*}[!t]
\centering
\includegraphics[width=0.7\linewidth]{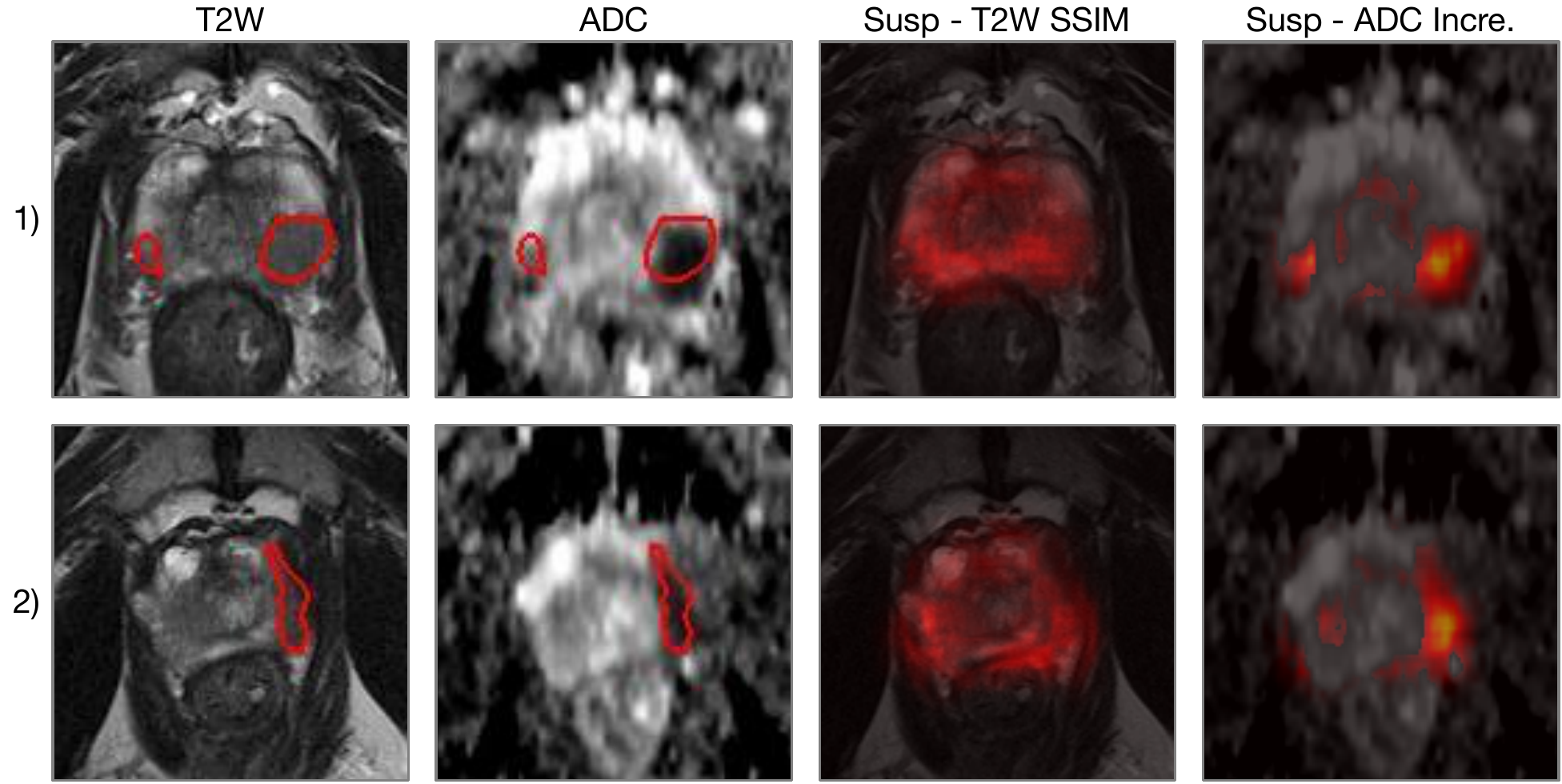}
\caption{The PCa suspiciousness maps with different distance functions for testing images. The red contours on T2w and ADC are the ground truth ROIs.}
\label{fig_distcomp}
\end{figure*}

\begin{figure*}[!t]
\centering
\includegraphics[width=0.7\linewidth]{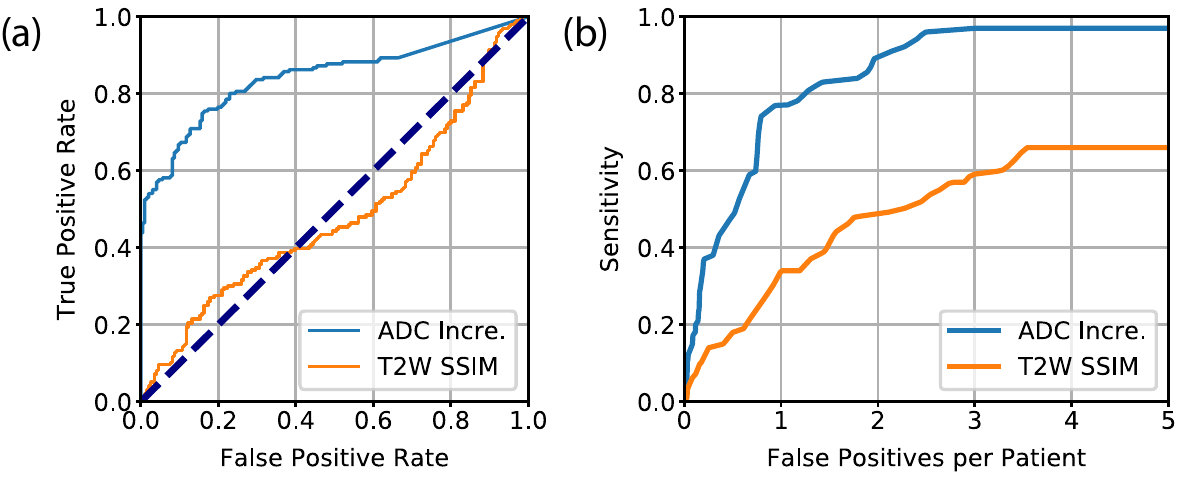} 
\caption{(a) ROC analysis for the classification between PCa lesions and non-PCa regions. (b) FROC analysis for lesion localization performance.}
\label{fig_ROC}
\end{figure*}

\subsection{Evaluation metrics}
\label{sec_evaluations}
The suspiciousness map by the baseline MRI model predicts pixel-wise levels of overall suspiciousness of prostate cancer and was used to distinguish between PCa and non-PCa regions~\cite{song2018computer,wang2018automated}. 
The PCa lesions were given by the ground truth ROIs, 
and non-PCa regions were defined as the same ground truth ROIs in the negative MRI testing cases. 
The average value over the region on the suspiciousness map is calculated as the predictive value for each ROI. 
The performance is evaluated by the receiver operating characteristic (ROC) analysis.

We also evaluated the lesion localization performance using the free-response receiver operating characteristic (FROC) analysis~\cite{litjens2014computer,wang2018automated}. 
The PCa localization points were determined by the local maximums of the suspiciousness map~\cite{wang2018automated}. 
A localization point was considered as a true positive if it is within 5mm of a ground truth lesion ROI, or it is otherwise a false negative~\cite{priester2017magnetic}. 
FROC measures the lesion detection sensitivity versus the average number of false positives for each patient. 

\subsection{Results}
\figurename~\ref{fig_distcomp} shows representative examples of the proposed cancer inference,
evaluated by an independent testing set. 
The red contours on both T2w and ADC are the ground truth annotations, 
indicating MRI-visible clinically significant PCa with histological confirmation. 
The overall pixel-wise suspiciousness with two distance functions, 
\textit{T2w SSIM} and \textit{ADC Increment},
are shown, and the \textit{ADC Increment} shows excellent visual predictability of PCa in both cases.

The ROC analysis for the classification between PCa lesions and non-PCa regions is shown in~\figurename~\ref{fig_ROC}. 
\textit{ADC Increment} (\textit{ADC Incre.}) achieved the area under the curve (AUC) of 0.84,
while the suspiciousness map using \textit{T2w SSIM} exhibited limited predictability for PCa.
Compared with ADC, T2w has a more diverse appearance for the normal tissues, 
potentially causing the suboptimal performance of the cancer inference.

The FROC analysis for lesion localization is shown in \figurename~\ref{fig_ROC}. 
\textit{ADC Increment} and \textit{T2w SSIM} had 77.0\% and 33.8\% detection sensitivity 
for PCa lesions with 1 false positive per patient, respectively, 
and 89.5\% and 48.8\% detection sensitivity at 2 false positives per patient. 
\textit{ADC Increment} received 95\% sensitivity at 2.44 false positives per patient, 
and \textit{T2w SSIM} reached its maximum sensitivity of 66.0\% at 3.54 false positives per patient.

\section{Discussion}
We proposed the novel cancer inference that can distinguish 
patients with and without PCa using weakly-supervised learning. 
We first identified 1,261 radiology-confirmed negative MRI cases out of all 3,127 in-house prostate MRI cases from 2016 to 2018.
The baseline MRI model was built to synthesize a partially obstructed MRI based on the understanding of the negative MRI appearance, and the cancer inference that predicts pixel-wise levels of overall suspiciousness was tested using a combination of negative and and highly curated positive MRI cases (n=232).
This weakly-supervised learning approach would be robust to any potential data bias 
due to the nature of the very-weak supervision and provides a scalable solution for training deep learning models.

The PCa detection sensitivity from previous studies ranged 
from 38.8\% to 89.8\% at 1 false positive per patient in the FROC analysis~\cite{litjens2014computer,tsehay2017biopsy,wang2018automated}. 
Despite the differences in dataset and inconsistencies of the lesion definition, 
our cancer inference via weakly supervised learning showed similar performance 
to the previously demonstrated models under strong supervision.
Compared with the fully-supervised methods trained with lesion annotations, 
the proposed method requires only negative MRI scans in training, 
which is a more practical and scalable approach to medical imaging 
since the method does not require a collection of large annotated prostate MRI data 
and is more suitable for multi-site, multi-vendor collaborations.

The regions are considered to be highly suspicious 
when the difference between the original and synthesized regions is large. 
We obtained the pixel-wise levels of overall suspiciousness by two distance functions,
\textit{T2w SSIM} and \textit{ADC Increment}. 
The primary reason to use these distance functions is that
T2w and ADC are typically used for anatomical and function imaging.
Future study could include the investigation of different distance functions, 
such as L2 norm and mutual information. 

\section{Conclusion}

We proposed the baseline MRI model via weakly supervised learning 
using a large collection of negative mp-MRI cases. 
The baseline MRI model was utilized to infer pixel-wise levels of overall suspiciousness of prostate cancer, without the need for using ground truth annotations.
The baseline MRI model was trained and validated using 1145 radiology-confirmed negative mp-MRI scans, and the cancer inference using the baseline MRI model was tested by a total of 232 independent mp-MRI scans. 
The proposed cancer inference via weakly supervised learning achieved an AUC of 0.84 in the ROC analysis 
and 77.0\% detection sensitivity at 1 false positive per patient in the FROC analysis  
using a separate dataset with histologically confirmed lesion annotations.

\section{Acknowledgments}

This work is supported by funds from the Integrated Diagnostics Program, Department of Radiological Sciences \& Pathology, David Geffen School of Medicine at UCLA.

{\small
\bibliographystyle{ieee}
\bibliography{egpaper_final}
}

\end{document}